\documentclass[%
 preprint,
 amsmath,amssymb,
 aps, physrev,
]{revtex4-2}

\usepackage{graphicx}
\usepackage{dcolumn}
\usepackage{bm}
\usepackage{amssymb}
\usepackage{amsmath}
\usepackage{physics}
\usepackage{float}
\usepackage{subcaption}
\usepackage{caption}
\usepackage{xcolor}
\usepackage{geometry}
\usepackage{overpic}
\usepackage[utf8]{inputenc}

\usepackage{hyperref}  
\usepackage{natbib}
\begin{document}

\preprint{APS/123-QED}

\title{\textbf{Network stochastic resonance under higher-order interactions} 
}%

\author{Zheng Wang}
\email{zhengwang@nuaa.edu.cn}
\affiliation{State Key Laboratory of Mechanics and Control for Aerospace Structures, College of Aerospace Engineering, Nanjing University of Aeronautics and Astronautics, Nanjing 210016, China}
 
\author{Jinjie Zhu}%
\email{Contact author: jinjiezhu@nuaa.edu.cn}
\affiliation{State Key Laboratory of Mechanics and Control for Aerospace Structures, College of Aerospace Engineering, Nanjing University of Aeronautics and Astronautics, Nanjing 210016, China}%

\author{Xianbin Liu}
\email{Contact author: xbliu@nuaa.edu.cn}
\affiliation{State Key Laboratory of Mechanics and Control for Aerospace Structures, College of Aerospace Engineering, Nanjing University of Aeronautics and Astronautics, Nanjing 210016, China}%

\date{\today}

\begin{abstract}
Although stochastic resonance phenomena are ubiquitous across various complex systems, the influence mechanisms of higher-order interactions remain elusive. Here, we address this gap by investigating stochastic resonance in coupled phase oscillators with triadic interactions on ring networks that feature periodic modulation of the triadic coupling strength and additive noise. Our analysis reveals that higher-order interactions create fundamentally different resonance landscapes through time-varying potential wells generated by periodic modulation of triadic coupling, enabling novel noise-enhanced processing mechanisms. Additionally, weaker pairwise coupling amplifies resonance effects, while moderate network connectivity appears optimal compared to extensive connections. Our findings establish fundamental principles for network stochastic resonance and provide insights for enhanced signal processing in complex networks.
\end{abstract}

\maketitle

\section{Introduction}

Stochastic resonance (SR)~\cite{sto1998,LINDNER2004,zhuweiqiu2006} represents a counterintuitive phenomenon whereby optimal noise levels amplify weak periodic signals rather than obscure them. This counterintuitive effect has been demonstrated in diverse systems including sensory neurons~\cite{MOSS2004,calim2021}, optomechanical systems~\cite{monifi2016,ricci2017}, electronic circuits~\cite{harmer2002}, aeroelastic airfoil systems~\cite{wang2025}, and biological networks~\cite{hanggi2002}. Applications encompass weak signal detection~\cite{chen2008,wu2024}, fault diagnosis in engineering systems~\cite{QIAO2017,LI2019}, and neural information processing~\cite{Guo2017,Guo2018}, etc.

Stochastic resonance phenomena in complex networks have been explored in various coupling topologies, from regular lattices to small-world networks~\cite{Gao2001,Li2024}. However, higher-order interactions are couplings that simultaneously connect three or more units in ways that resist decomposition into pairwise terms, fundamentally altering collective network dynamics~\cite{battiston2021,Bianconi2021,Bick2023,PRE2022}. These nonlinear couplings prove essential for understanding complex processes, including social contagion~\cite{sct2013,PhysRevLe2021}, neural computation~\cite{petri2014,santoro2024}, and oscillator synchronization~\cite{matthew2019,Gengel2020}. While numerous studies have examined their role in synchronization and stability phenomena~\cite{Mill2020,zhangyz2024}, their influence on stochastic resonance remains underexplored.

Given that SR depends critically on the structure of dynamical landscapes and the accessibility of different system states, higher-order interactions that can fundamentally restructure both the number and stability properties of attracting states present compelling opportunities for novel resonance mechanisms~\cite{Skardal2020,Skardal2023,zhangyz2024,wangz2025} . This gap motivates our investigation into how triadic couplings modify classical resonance phenomena in oscillatory networks.

In this work, we investigate network stochastic resonance in coupled oscillator networks with triadic interactions, developing a theoretical framework that incorporates both pairwise and triadic connections, along with periodic modulation of the triadic coupling strength and additive noise. Using the signal-to-noise ratio and mean response amplitude as complementary metrics, we characterize the parameter dependencies and examine how variations in network structure influence the manifestation of resonance.

\section{Results}

\subsection{Model framework}

Higher-order interactions in oscillatory networks provide a rich framework for investigating stochastic resonance phenomena that transcend conventional pairwise coupling paradigms. Our theoretical approach builds upon phase reduction methodology~\cite{kuramoto1984,zhangyz2024} to develop a generalized Kuramoto model~\cite{STROGATZ2000} with both pairwise and triadic interactions among identical oscillators. The mathematical foundation of our system emerges from the general dynamical framework~\cite{leon2025}:

\begin{equation}
\dot{\theta}_j = \omega_j + \varepsilon \sum_{k,l=1}^{N} A_{jkl}\Gamma_{jkl}(\Delta\theta_{kj}, \Delta\theta_{lj}),
\end{equation}
in this formulation, $\theta_j$ captures the instantaneous phase of the $j$-th oscillator, with $\omega_j$ representing its natural frequency. The network topology manifests through the tensor $A_{jkl}$, which equals one for valid triadic configurations involving distinct nodes $j$, $k$, and $l$, and zero otherwise. The parameter $\varepsilon$ scales the overall coupling intensity, while $\Gamma_{jkl}(\Delta\theta_{kj}, \Delta\theta_{lj})$ encodes the interaction kernel dependent on phase differences.

We restrict our analysis to ring topologies with finite coupling range $r$ to maintain computational feasibility. Within this configuration, the deterministic system evolves according to~\cite{zhangyz2024}:

\begin{equation}
\dot{\theta}_i = \frac{\sigma}{2r} \sum_{j=i-r}^{i+r} \sin(\theta_j - \theta_i) + \frac{\sigma_\Delta}{2r(2r-1)} \sum_{j=i-r}^{i+r} \sum_{k=i-r}^{i+r} \sin(\theta_j + \theta_k - 2\theta_i),
\end{equation}
where $i$, $j$, and $k$ are distinct ensures that triadic couplings involve three different oscillators. Here, $\sigma$ and $\sigma_\Delta$ represent the strengths of pairwise and triadic interactions, respectively. The denominators $2r$ and $2r(2r-1)$ provide proper normalization based on connection counts. We employ a co-rotating frame where $\omega = 0$ for uniform intrinsic frequencies. Following previous studies~\cite{zyz2021,zhangyz2024,wangz2025}, our computational studies utilize $n = 83$ units with $r = 2$.

To explore stochastic resonance mechanisms, we incorporate periodic modulation into the triadic coupling strength and additive stochastic fluctuations into the system dynamics. This kind of time-varying coupling approach has been extensively studied in coupled oscillator networks~\cite{cumin2007,petkoski2012}. The system evolves under the influence of time-dependent coupling:

\begin{equation}
\sigma_\Delta(t) = \sigma_{\Delta,0} + A \sin(2\pi f t),
\end{equation}
where $\sigma_{\Delta,0} = 5$ represents the baseline triadic coupling strength, $A$ controls the forcing amplitude, and $f$ denotes the driving frequency. Additionally, each oscillator experiences independent Gaussian white noise $\xi_i(t)$ with zero mean and intensity $D$, such that $\langle \xi_i(t) \xi_j(t') \rangle = 2D \delta_{ij} \delta(t-t')$. Then, the modified dynamical equations become:
\begin{equation}
\dot{\theta}_i = \frac{\sigma}{2r} \sum_{j=i-r}^{i+r} \sin(\theta_j - \theta_i) + \frac{\sigma_\Delta(t)}{2r(2r-1)} \sum_{j=i-r}^{i+r} \sum_{k=i-r}^{i+r} \sin(\theta_j + \theta_k - 2\theta_i) + \xi_i(t).
\end{equation}

This framework allows us to investigate how the interplay between periodic driving and noise affects the emergence of collective synchronization patterns. The degree of local synchronization between each oscillator and its nearest neighbors is quantified by the local order parameter~\cite{zhangyz2024}:
\begin{equation}
O_j = \frac{1}{2r+1} \sum_{k=j-r}^{j+r} e^{i\theta_k},
\end{equation}
for $j = 1, ..., n$. An oscillator $j$ is designated as ordered when $|O_j| \geq 0.85$, and as disordered when $|O_j| < 0.85$. To quantify collective behavior, we introduce the order parameter:
\begin{equation}
R = \frac{n_{\text{order}}}{n},
\end{equation}
which represents the fraction of ordered oscillators within the network, where $n_{\text{order}}$ is the total number of oscillators satisfying $|O_j| \geq 0.85$. This threshold selection ensures that $R$ approaches zero under conditions of strong triadic coupling, aligning with direct observation of the system states. Although this criterion is empirically motivated, ordered configurations typically demonstrate $|O_j| \approx 1$ for systems with modest winding numbers and limited coupling ranges, both characteristics present in our network setup.

\subsection{Network stochastic resonance characterization}

This order parameter $R$ serves as our primary diagnostic tool, enabling us to assess the extent of collective synchronization while monitoring how the combination of triadic couplings, periodic forcing, and noise influences group dynamics. More importantly, it allows us to identify signatures of stochastic resonance behavior and characterize the complex interplay between random fluctuations and deterministic evolution. The parameter spans from complete order ($R = 1$) to complete disorder ($R = 0$), with intermediate values capturing chimera states and other spatially heterogeneous configurations.

\begin{figure}[htbp]
\centering
\captionsetup{justification=raggedright, singlelinecheck=false}
\includegraphics[width=1\textwidth]{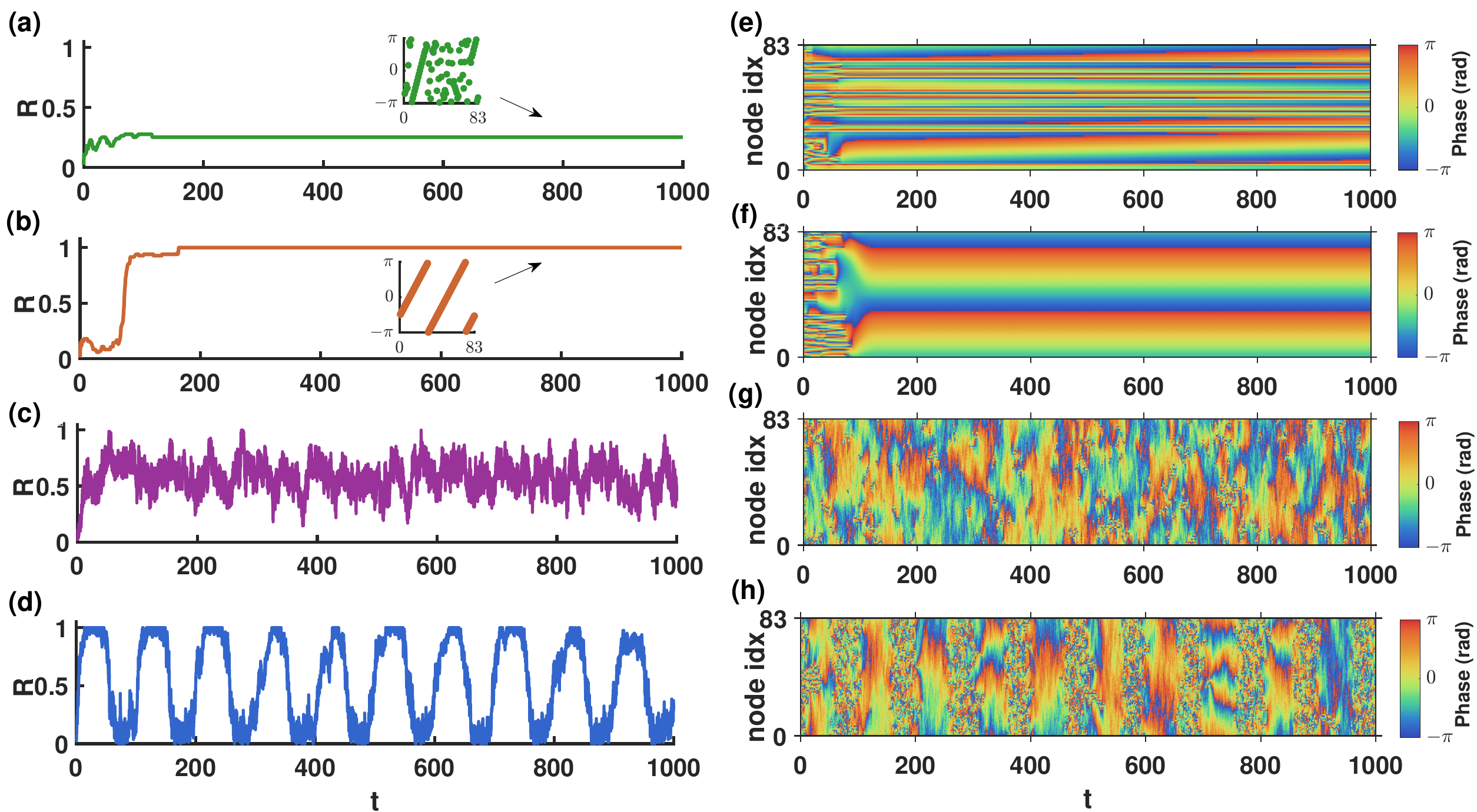}
\caption{Time evolution of the order parameter $R$ under different noise and modulation conditions: (a) $D=0$, $A=0$; (b) $D=0$, $A=4.5$; (c) $D=1$, $A=0$; (d) $D=1$, $A=4.5$. Panels (e-h) show the corresponding spatiotemporal patterns of oscillator phases for conditions (a-d) respectively. Parameters: $n = 83$, $r = 2$, $\sigma = 1$, $\sigma_{\Delta,0} = 5$, $f = 0.01$.}
\label{fig:1}
\end{figure}

FIG.~\ref{fig:1} illustrates the distinct dynamical behaviors emerging under various combinations of noise intensity $D$ and forcing amplitude $A$. In panel (a), the pure deterministic system ($D=0$, $A=0$) maintains the order parameter at an intermediate value, indicating that the system settles into a chimera state without external perturbations. Under the triadic coupling strength $\sigma_{\Delta,0} = 5$, chimera and disordered states dominate the basin of attraction~\cite{zhangyz2024}. The corresponding spatiotemporal pattern in panel (e) reveals the characteristic phase distribution of a chimera state. Panel (b) demonstrates the deterministic response to periodic forcing ($D=0$, $A=4.5$), where the periodic signal drives the system from the non-ordered attractor basin into the ordered state basin. Once the system stabilizes in the ordered state, it remains there regardless of the continuing periodic variations, as evidenced by the sustained high value of the order parameter and the uniform phase pattern shown in panel (f). Panel (c) shows pure noise effects ($D=1$, $A=0$) leading to irregular fluctuations around intermediate order parameter values, with the spatiotemporal dynamics in panel (g) displaying random phase transitions. Most significantly, panel (d) reveals the stochastic resonance phenomenon where both noise and periodic forcing are present ($D=1$, $A=4.5$). Here, the order parameter displays enhanced oscillations, demonstrating the classical signature of stochastic resonance where optimal noise intensity amplifies the system's response to weak periodic signals. The spatiotemporal pattern in panel (h) shows coherent periodic switching between ordered and disordered phases. The spatiotemporal patterns in panels (e-h) reveal the characteristic phase distributions for each condition, with the complete temporal evolution of the system dynamics illustrated in Supplementary Video 1.

It is important to distinguish this network-level stochastic resonance from traditional single-system stochastic resonance. Here, we define network stochastic resonance as the optimal enhancement of collective synchronization (measured by the order parameter $R$) rather than individual system responses. The resonance phenomenon emerges at the collective level through the interplay between noise-induced transitions and the time-varying energy landscape created by periodic triadic coupling modulation, which differs fundamentally from individual resonance effects that may occur without global network coherence.

The emergence of this stochastic resonance behavior can be understood through the interplay between the triadic coupling modulation and the underlying energy landscape. According to previous studies~\cite{wangz2025}, as the triadic coupling strength $\sigma_\Delta$ increases, the potential wells associated with ordered states progressively deepen, while within a certain range, the relative potential depth of disordered states remains significantly shallower compared to ordered states. The periodic modulation of $\sigma_\Delta$ thus creates a time-varying energy landscape where the potential wells alternately become shallow and deep. As illustrated in FIG.~\ref{fig:1}(d), when $\sigma_\Delta$ is small, the ordered states cannot withstand noise perturbations effectively due to their shallow potential wells, causing noise to prevent the system from rapidly returning to ordered configurations, which results in low values of the order parameter. Conversely, when $\sigma_\Delta$ becomes large, the deepened potential wells of ordered states provide sufficient stability against noise-induced transitions, preventing the system from being pushed toward disordered states. The periodic modulation of $\sigma_\Delta$ thus creates a time-varying energy landscape where the potential wells alternately become shallow and deep. During phases when $\sigma_\Delta$ is small and potential wells are shallow, moderate noise can effectively disrupt the ordered state, while during phases when $\sigma_\Delta$ is large and potential wells are deep, the same noise level is insufficient to destabilize the system. This alternating dynamics, when matched with appropriate noise intensity, enables the system to exhibit enhanced periodic response, manifesting as the characteristic stochastic resonance phenomenon.

\subsection{Effects of noise intensity}

To quantify the stochastic resonance phenomenon in our oscillatory network, we utilize two complementary metrics (detailed in Methods~\ref{methods}) that provide cross-validation for the accuracy of our results while capturing distinct aspects of the system's response to periodic forcing. Both metrics undergo ensemble averaging over $M$ independent realizations to ensure statistical reliability. The consistency between these two independent measures provides confidence in our characterization of the stochastic resonance phenomenon.

\begin{figure}[htbp]
\centering
\captionsetup{justification=raggedright, singlelinecheck=false}
\includegraphics[width=0.9\textwidth]{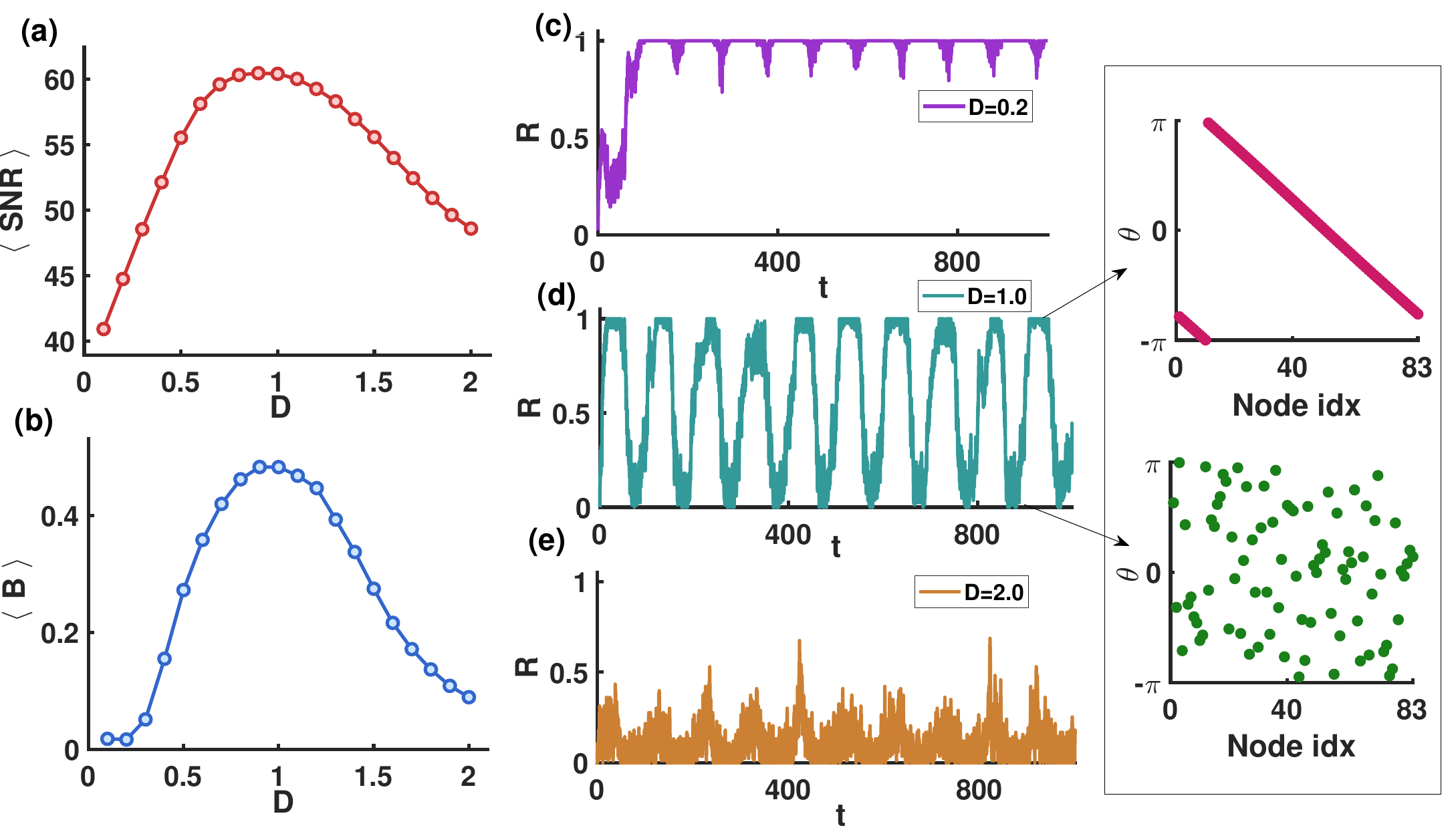}
\caption{Stochastic resonance indicators as functions of noise intensity $D$. (a) Signal-to-noise ratio $\langle \text{SNR} \rangle$. (b) Mean response amplitude $\langle B \rangle$. (c-e) Time series of order parameter $R$ for $D = 0.2, 1, 2.0$, respectively. Right panels show phase configurations of oscillators after removing noise at the arrowed time instants. Parameters: $n = 83$, $r = 2$, $\sigma = 1$, $\sigma_{\Delta,0} = 5$, $A = 4.5$, $f = 0.01$.}
\label{fig:2}
\end{figure}

Having demonstrated the occurrence of stochastic resonance in our system, we now systematically analyze how noise influences the system's response to periodic forcing, beginning with the fundamental relationship between noise intensity and signal amplification. FIG.~\ref{fig:2} demonstrates the fundamental characteristics of stochastic resonance through the behavior of our two complementary indicators as functions of noise intensity $D$. Panel (a) shows that the signal-to-noise ratio exhibits a pronounced increase with increasing noise strength, rising sharply from approximately 40 at $D = 0.1$ to nearly 60 at $D = 1$, after which it reaches saturation. At low noise levels, insufficient noise prevents effective state switching, resulting in weak signal detection. As noise intensity increases beyond $D \approx 1$, the SNR begins to decline gradually, indicating that further noise addition does not significantly improve the signal detection capability. Panel (b) presents the mean response amplitude $\langle B \rangle$, which exhibits the characteristic bell-shaped curve that defines classical stochastic resonance. The amplitude increases from near-zero values at low noise intensities, reaches a maximum of approximately 0.48 at $D \approx 1$, and subsequently decreases for higher noise levels. This non-monotonic dependence represents the hallmark of stochastic resonance: optimal noise levels enhance the system's response to weak periodic signals, while excessive noise degrades the coherent response. The peak location at $D \approx 1$ identifies the optimal noise intensity that maximizes the oscillatory response amplitude, providing a clear quantitative measure of the resonance condition. Panels (c-e) illustrate the temporal evolution of the order parameter $R$ at three representative noise intensities: $D=0.2$ (weak noise), $D=1$ (optimal noise), and $D=2.0$ (strong noise), corresponding to different regimes of the resonance curve. The accompanying phase snapshots on the right reveal the underlying synchronization mechanisms: weak noise is insufficient to drive the system out of ordered potential wells, resulting in predominantly stable synchronized states; optimal noise enables effective transitions between ordered and disordered configurations, producing coherent oscillations that follow the periodic modulation; while strong noise overwhelms the underlying dynamics, masking the periodic signal and leading to largely irregular fluctuations.

\subsection{Effects of parameters}

\begin{figure}[htbp]
\centering
\captionsetup{justification=raggedright, singlelinecheck=false}
\includegraphics[width=0.95\textwidth]{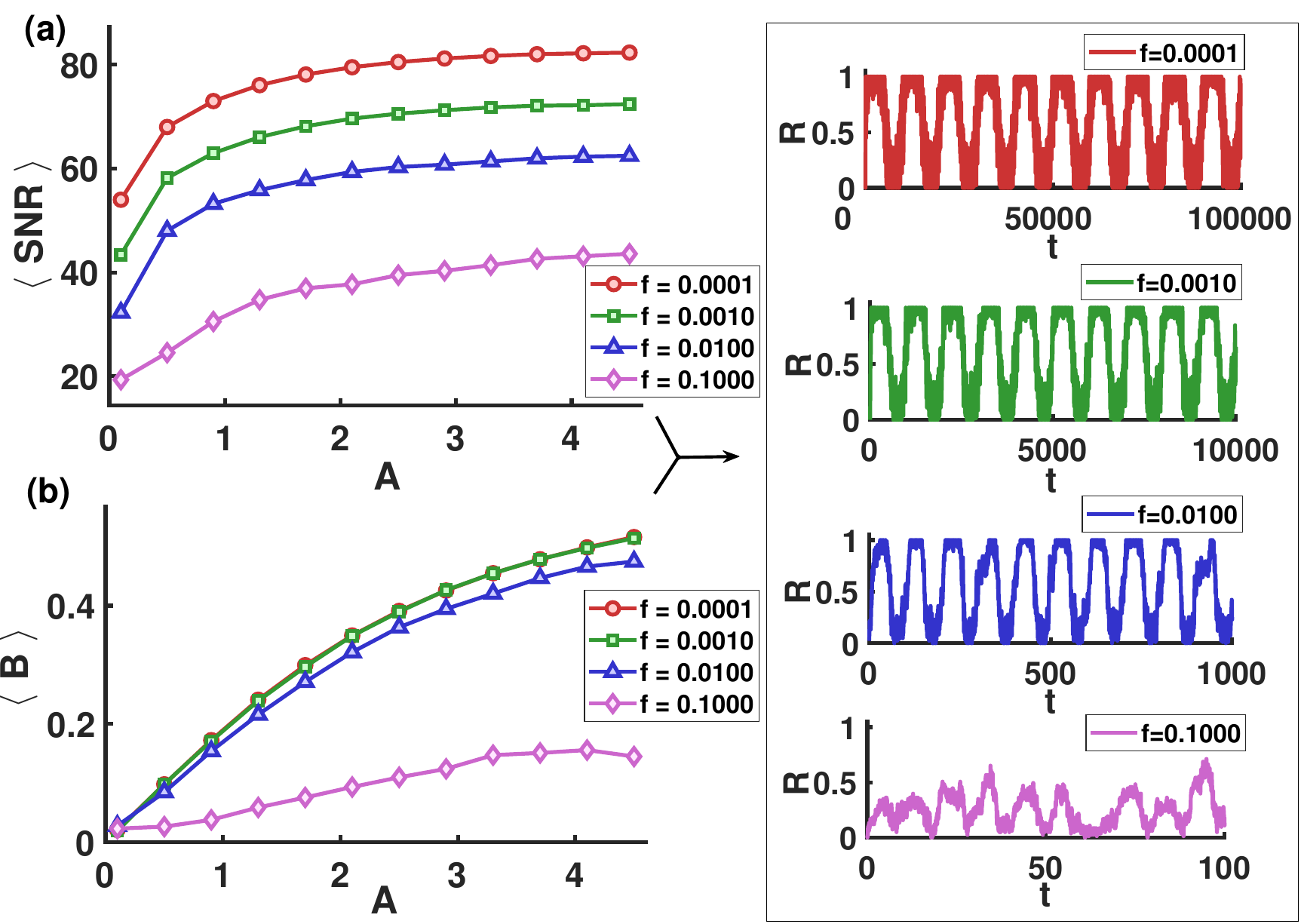}
\caption{Stochastic resonance indicators as functions of driving amplitude $A$ for different frequencies $f$. (a) Signal-to-noise ratio $\langle \text{SNR} \rangle$. (b) Mean response amplitude $\langle B \rangle$. Right panels show representative time series of order parameter $R$ for different frequencies at $A = 4.5$. Parameters: $n = 83$, $r = 2$, $\sigma = 1$, $\sigma_{\Delta,0} = 5$, $D = 1$.}
\label{fig:3}
\end{figure}

Having confirmed the existence of stochastic resonance, we next examine how the phenomenon depends on the characteristics of the periodic driving signal itself. FIG.~\ref{fig:3} explores how the stochastic resonance indicators respond to variations in driving amplitude $A$ across different frequencies. Panels (a) and (b) demonstrate that both the SNR and mean response amplitude $\langle B \rangle$ increase monotonically with forcing strength for all tested frequencies, which is expected since stronger driving provides more pronounced perturbations that can more effectively induce transitions between different dynamical states. However, the system exhibits a clear frequency performance ranking: progressively lower frequencies achieve increasingly higher resonance indicators, demonstrating enhanced sensitivity as the driving frequency decreases. The underlying mechanism can be understood through the time-scale matching perspective: the system's intrinsic dynamics, governed by the interplay between triadic coupling modulation and noise-induced transitions, operate on characteristic time scales that are better matched to slower driving frequencies. The right panels show representative time series of the order parameter $R$ at $A = 4.5$ for different frequencies, clearly illustrating how progressively slower frequencies produce more coherent and pronounced oscillatory responses, while faster frequencies result in increasingly irregular dynamics that poorly follow the periodic modulation. Since stochastic resonance becomes quite pronounced when the frequency reaches $f=0.01$, we use $f=0.01$ as the standard frequency for subsequent investigations.

\subsection{Network topology effects}

Having established the fundamental characteristics of stochastic resonance in our coupled oscillator network with specific coupling parameters, we now investigate how variations in the network structure itself influence the resonance phenomenon. Specifically, we examine the effects of different pairwise coupling strengths $\sigma$ and coupling ranges $r$ on the stochastic resonance indicators while maintaining the optimal conditions identified in the previous analysis ($D = 1$, $A = 4.5$, $f = 0.01$).

\begin{figure}[htbp]
\centering
\captionsetup{justification=raggedright, singlelinecheck=false}
\includegraphics[width=0.95\textwidth]{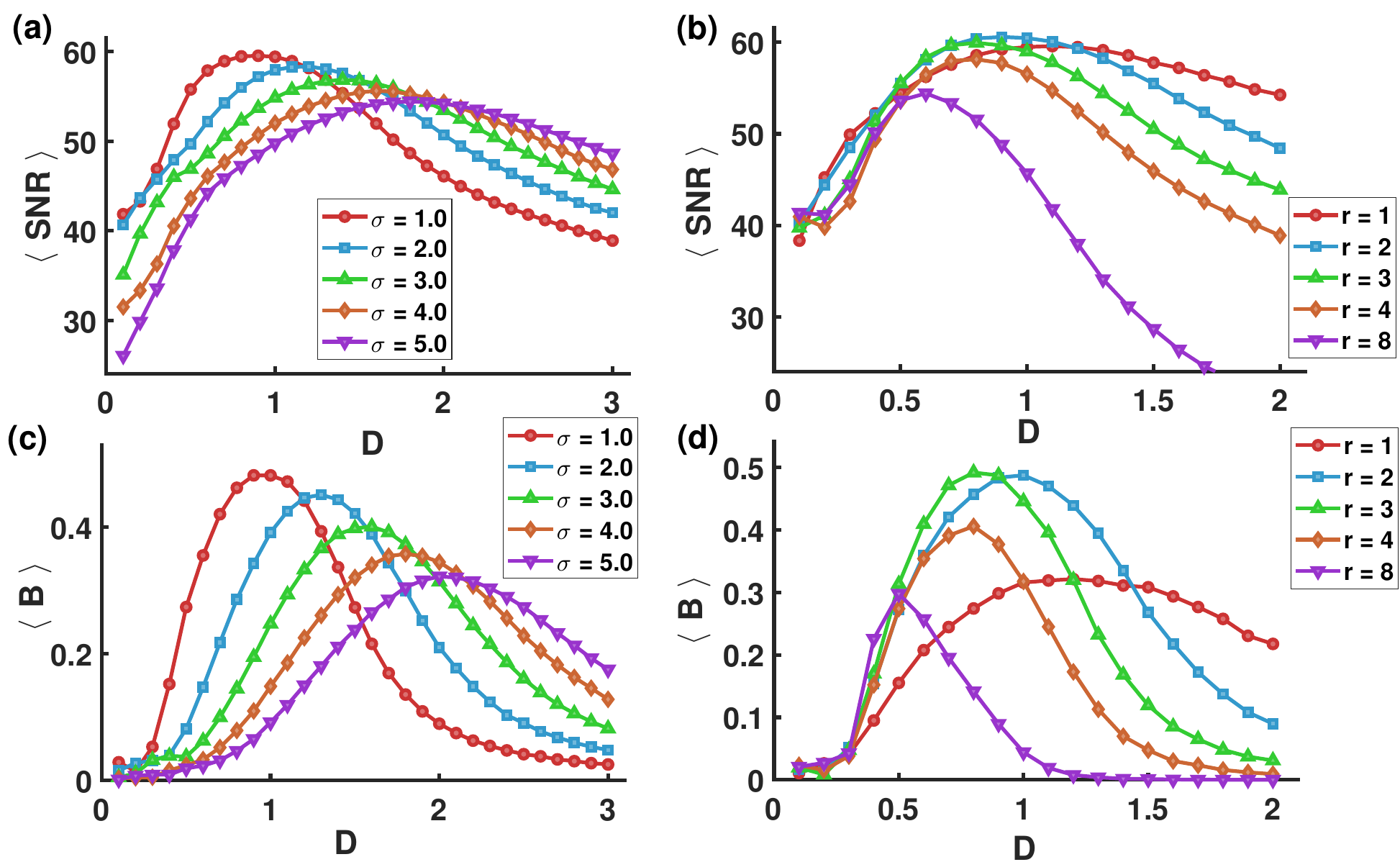}
\caption{Stochastic resonance indicators under different coupling structures. (a, c) Signal-to-noise ratio $\langle \text{SNR} \rangle$ and mean response amplitude $\langle B \rangle$ as functions of noise intensity $D$ for various pairwise coupling strengths $\sigma$. (b, d) The same indicators for different coupling ranges $r$. Parameters: $D = 1$, $A = 4.5$, $f = 0.01$, $n = 83$, $\sigma_{\Delta,0} = 5$. For panels (a, c): $r = 2$. For panels (b, d): $\sigma = 1$.}
\label{fig:4}
\end{figure}

FIG.~\ref{fig:4} presents a comprehensive analysis of how coupling structure modifications affect the stochastic resonance behavior. Panels (a) and (c) explore the influence of pairwise coupling strength $\sigma$ on both indicators. The results reveal a systematic trend: as the pairwise coupling strength increases from $\sigma = 1.0$ to $\sigma = 5.0$, the characteristic bell-shaped curves become progressively flattened, with peak values decreasing and optimal noise intensities shifting toward higher values. Both SNR and mean response amplitude exhibit the most pronounced stochastic resonance effects at the weakest coupling strength, with the resonance phenomenon becoming increasingly suppressed as coupling strength increases. This behavior suggests that stronger pairwise coupling competes with the triadic interactions that drive the stochastic resonance phenomenon, effectively suppressing the noise-enhanced signal detection capability.

The physical mechanism underlying this suppression can be understood through the competition between different coupling pathways and the fundamental nature of higher-order interactions. Strong pairwise coupling tends to create more rigid local synchronization patterns that resist the noise-induced transitions necessary for stochastic resonance. When $\sigma$ is large, the system becomes dominated by nearest-neighbor interactions, reducing the effectiveness of the higher-order triadic couplings that provide the nonlinear dynamics essential for resonance behavior.

Panels (b) and (d) examine the role of coupling range $r$ in determining stochastic resonance characteristics. The coupling range fundamentally alters the network connectivity, with larger $r$ values creating more densely connected topologies. For the smallest coupling range ($r = 1$), both indicators exhibit relatively weak stochastic resonance effects. Interestingly, as the coupling range increases to $r = 2$, we observe a notable enhancement in resonance performance, suggesting the existence of an optimal intermediate connectivity regime. However, further increases in coupling range ($r = 3, 4, 8$) lead to progressively weaker resonance effects, with a consistent trend where the optimal noise intensity progressively shifts toward lower values while the peak amplitudes generally decrease.

The non-monotonic enhancement from $r = 1$ to $r = 2$ can be understood through a critical mechanism of energy landscape restructuring. At $r = 1$, the limited nearest-neighbor connectivity constrains the effective scope of triadic interactions. While triadic coupling does enhance the depth of disordered state potential wells~\cite{wangz2025}, this enhancement remains too shallow relative to ordered state wells due to the restricted coupling range, failing to establish effective dynamical competition. Consequently, the system is predominantly dominated by ordered states, lacking the sufficient state-switching capacity necessary for stochastic resonance. The transition to $r = 2$ enables triadic interactions to operate more fully through expanded next-nearest-neighbor connections, creating disordered state potential wells comparable in depth to ordered states, thereby establishing the balanced energy landscape that supports optimal noise-induced transitions. As the coupling range increases further, the system transitions toward over-connectivity. Large coupling ranges promote global coupling behaviors that tend to homogenize the system's response, reducing sensitivity to the optimal noise levels that characterize stochastic resonance. Moreover, increased connectivity facilitates the emergence of chimera states and completely disordered states. As the coupling range increases, these competing states develop larger basins of attraction and deeper potential wells compared to ordered twisted states~\cite{zhangyz2024}, making the system more likely to become trapped in dynamical regimes that exhibit weaker stochastic resonance responses to periodic driving.

These findings demonstrate that the network structure plays a crucial role in determining the strength and characteristics of stochastic resonance in coupled oscillator systems. The optimal resonance conditions emerge from a delicate balance between pairwise and triadic interactions, with moderate coupling strengths and limited connectivity ranges providing the most favorable conditions for noise-enhanced signal detection.

\section{Discussion}

In this work, we have comprehensively investigated network stochastic resonance phenomena in coupled oscillator networks with higher-order interactions, revealing fundamental mechanisms that govern noise-enhanced signal detection in complex dynamical systems. Our study demonstrates that triadic couplings, when combined with periodic forcing and optimal noise levels, can produce robust stochastic resonance effects that significantly enhance the system's ability to detect and amplify weak periodic signals.

Through systematic analysis using two complementary indicators, the signal-to-noise ratio and mean response amplitude, we have established that stochastic resonance emerges from the delicate interplay between periodic modulation of triadic coupling strength and the underlying energy landscape. The periodic variation of $\sigma_\Delta(t)$ creates a time-varying energy landscape where potential wells alternately become shallow and deep, enabling optimal noise levels to synchronize with the driving frequency and facilitate coherent state transitions. This mechanism represents a novel pathway for achieving noise-enhanced signal processing in higher-order networked systems.

Our investigation of parameter dependencies reveals several key insights into the optimization of stochastic resonance. The noise intensity dependence exhibits the characteristic bell-shaped curve for mean response amplitude, with optimal performance occurring at $D \approx 1$, while the signal-to-noise ratio shows a monotonic increase followed by a gradual decline. The forcing amplitude and frequency dependencies demonstrate that stronger driving enhances resonance effects, while lower frequencies prove more effective in eliciting coherent network responses. These findings provide practical guidelines for tuning system parameters to achieve optimal noise-enhanced signal detection.

The exploration of different coupling structures reveals the critical role of network topology in determining stochastic resonance characteristics. Weaker pairwise coupling ($\sigma = 1.0$) produces stronger resonance effects, as excessive pairwise interactions compete with triadic couplings and create rigid synchronization that inhibits necessary noise-induced transitions. Similarly, moderate coupling ranges ($r = 2$) prove optimal, where triadic interactions can establish effective dynamical competition between states, contrasting with insufficient triadic effects at $r = 1$ and over-connectivity at $r = 8$ that suppress resonance.

Our results contribute to the broader understanding of collective dynamics in higher-order networks and provide insights relevant to diverse applications ranging from neural information processing to engineered oscillator arrays. The demonstration that moderate network connectivity and balanced coupling strengths optimize stochastic resonance performance offers design principles for systems requiring enhanced sensitivity to weak signals. Furthermore, the complementary nature of our two indicators provides a robust framework for characterizing stochastic resonance that can be applied to other complex dynamical systems.

Future work could extend these findings by exploring different higher-order coupling functions~\cite{Park2024}, investigating the role of network heterogeneity~\cite{Per2016,Javier2025}, and examining stochastic resonance in adaptive networks~\cite{Ha2016,Papadopoulos2017} where coupling strengths evolve dynamically. Additionally, experimental validation of these theoretical predictions in physical oscillator networks would provide valuable confirmation of the mechanisms identified in this study. Electrochemical oscillator arrays, which have successfully demonstrated phase model predictions for synchronization phenomena~\cite{kiss2005,jorge2025}, offer a promising experimental platform for implementing the triadic coupling schemes and periodic modulation required to observe network stochastic resonance. The theoretical framework established in this work provides new insights into how higher-order interactions can fundamentally alter noise-enhanced signal processing in networked systems, offering design principles for optimizing collective stochastic resonance that could guide future experimental implementations and engineering applications.
\section{\label{methods}Methods}

\subsection{Signal-to-noise ratio}

The first indicator employs a signal-to-noise ratio (SNR) based on power spectral density analysis. Given the order parameter time series $\{R(t_k)\}_{k=0}^{N-1}$ with $t_k = k \Delta t$, where $\Delta t$ represents the sampling interval and $N$ is the total number of data points, we perform discrete Fourier transformation:

\begin{equation}
X_m = \sum_{k=0}^{N-1} R(t_k) e^{-2\pi i mk/N},
\end{equation}
where $m$ is the frequency index ranging from $0$ to $N-1$. The corresponding frequency grid is defined as $f_m = m/(N \Delta t)$ with frequency resolution $\Delta f = 1/T$, where $T = N \Delta t$ is the total sampling duration.

The single-sided power spectral density is computed as $S(f_m) = |X_m|^2/N$. Signal power is extracted at the driving frequency $f_0$ by identifying the closest frequency grid point:

\begin{equation}
P_{\text{signal}} = S(f_{\text{signal}}), \quad \text{where} \quad m_{\text{signal}} = \arg\min_m |f_m - f_0|.
\end{equation}
Noise power is estimated as the background spectral density at the driving frequency:

\begin{equation}
P_{\text{noise}} = S_N(f_0),
\end{equation}
where $S_N(f_0)$ represents the noise background level at frequency $f_0$. The linear SNR is then calculated as:

\begin{equation}
\text{SNR} = \frac{P_{\text{signal}}}{P_{\text{noise}}}.
\end{equation}

\subsection{Mean response amplitude}

Our second indicator focuses on the mean response amplitude, which provides an alternative perspective on the system's dynamical behavior. Following the approach of extracting the amplitude of periodic responses, we decompose the order parameter time series into sine and cosine components:

\begin{align}
B_S &= \frac{2}{T} \int_0^T R(t) \sin(2\pi f_0 t) dt, \\
B_C &= \frac{2}{T} \int_0^T R(t) \cos(2\pi f_0 t) dt,
\end{align}
where $f_0$ is the driving frequency and $T$ is the total observation time. The mean response amplitude is then obtained as:

\begin{equation}
B = \sqrt{B_S^2 + B_C^2}.
\end{equation}

This quantity directly measures the magnitude of the system's oscillatory response at the driving frequency, complementing the frequency-domain analysis provided by the SNR metric. Under stochastic resonance conditions, both indicators typically exhibit a characteristic non-monotonic dependence on noise intensity, with optimal values occurring at intermediate noise levels. Both metrics undergo ensemble averaging over $M$ independent realizations to ensure statistical reliability:

\begin{align}
\langle \text{SNR} \rangle &= \frac{1}{M} \sum_{m=1}^{M} \text{SNR}^{(m)}, \\
\langle B \rangle &= \frac{1}{M} \sum_{m=1}^{M}  B ^{(m)},
\end{align}
where $M$ is the number of independent simulation runs and the superscript $(m)$ denotes the $m$-th realization. The consistency between these two independent measures provides confidence in our characterization of the stochastic resonance phenomenon.

\begin{acknowledgments}
JZ acknowledges the support from the National Natural Science Foundation of China (Grant Nos. 12202195 and 12572037). XL thanks the National Natural Science Foundation of China (Grant No. 12172167) for financial support.
\end{acknowledgments}

\section*{Author contributions}
Z.W.: Conceptualization, Methodology, Software, Formal analysis, Investigation, Data curation, Writing: original draft, Visualization. J.Z.: Conceptualization, Methodology, Writing: review and editing, Supervision, Project administration, Funding acquisition. X.L.: Conceptualization, Methodology, Writing: review and editing, Supervision, Project administration, Funding acquisition.

\section*{Competing interests}
The authors declare no competing interests.

\section*{Supplementary material}
\textbf{Supplementary Movie.} Dynamic evolution of network stochastic resonance.

\section*{Data availability}
All data needed to evaluate the conclusions are present in the paper. Additional data related to this paper may be requested from the authors.

\section*{Code availability}
The codes that support the findings of this study are available from the corresponding authors upon reasonable request.


%

\end{document}